# Superfluorescent emission in electrically pumped semiconductor laser

D. L. Boiko[1], X. Zeng[1], T. Stadelmann[1], S. Grossmann[1], A. Hoogerwerf[1], T. Weig[2], U. T. Schwarz[2], L. Sulmoni[3], J.-M. Lamy[3], N Grandjean[3]

[1]Centre Suisse d'Électronique et de Microtechnique SA (CSEM), CH-2002 Neuchâtel, Switzerland

[2] Fraunhofer Institut für Angewandte Festkörperphysik IAF, 79108 Freiburg, Germany

[3] Institute of Condensed Matter Physics (ICMP), École Polytechnique Fédérale de Lausanne (EPFL), CH-1015 Lausanne, Switzerland

Abstract:

We report superfluorescent (SF) emission in electrically pumped InGaN/InGaN QW lasers with saturable absorber. In particular, we observe a superlinear growth of the peak power of SF pulses with increasing amplitude of injected current pulses and attribute it to cooperative pairing of electron–hole (e-h) radiative recombinations. The phase transitions from amplified spontaneous emission to superfluorescence and then to lasing regime is confirmed by observing (i) abrupt peak power growth accompanied by spectral broadening, (ii) spectral shape with hyperbolic secant envelope and (iii) red shift of central wavelength of SF emission pulse. The observed red shift of SF emission is shown to be caused by the pairing of e-h pairs in an indirect cooperative X-transition.

Key words: superfluorescence, superradiance, Cooper pairs

Corresponding author: Dmitri Boiko , dmitri.boiko@csem.ch

Dicke superradiance [1] or superfluorescence (SF) [2] with optical pulse peak power proportional to the square of the number of inverted atoms ($I_{peak} \propto N^2$) is widely studied in atomic (and molecular) gases [3], offering homogeneous line broadening and long dephasing time $T_2$. SF in solids is more challenging because of the short dephasing time. The pulse width $\tau_p \propto N^{-1}$ is inversely proportional to the growth rate (increment) of the field. There were several claims on SR emission in optically pumped solids, even one demonstrating superliner peak power growth [4]. This includes the most detailed study of SF in crystalline $KCl:O_2^-$ where transitions between levels in super oxide ion $O_2^-$ suffers from strong inhomogeneous broadening $\sim T_2^{-1}$ [5]. Realization of SF in semiconductors is a very controversial subject. Here, in addition to strong inhomogeneous broadening, the e-h pairs exhibit a strong dispersion of energy within the bands. According to [6], which is largely based on results from Ref. [7] regarding the requirements of dephasing time $T_2$ and so-called cooperative frequency $w_c$, no SF is possible in a bulk semiconductor material because of the presence of inhomogeneous broadening and dispersion. One needs to create a singularity in the density of states (DOS) spectrum, as in case of atomic bound electrons in [1] or [7]. This can be achieved by applying a magnetic field to electron-hole (e-h) plasma in a QW or by using QDs or exciton transition. Magnetic field in bulk materials or uniform QWs do not provide DOS singularity. Moreover, inherent inhomogeneous broadening caused by QW thickness variation smoothen the DOS distribution and prevent experimental realization of SF in semiconductor QWs in the absence of applied magnetic field. Observation of SF from optically excited excitons in ZnO QDs [8] and magneto-excitons in InGaAs/GaAs quantum wells placed in strong magnetic field [4] have been confirmed by demonstrating superlinear growth of the peak power $I_{peak} \propto N^{2.4}$ and $I_{peak} \propto N^{3/2}$, respectively. When the coherent pump laser wavelength is tuned above the band edge, one should expect that intraband relaxation should result in a completely decoherent initial state of the system before SF. Nevertheless, a demonstration of SF under electrical injection of carriers (incoherent pumping) would provide definitive evidence. Indeed, observation of SF in [4] might be questioned as explanation for observed $I_{peak} \propto N^{3/2}$ growth instead of $I_{peak} \propto N^2$ is based on [7], in which the growth rate (increment) of the field does not account for the depletion of the excited atoms and predicts $\tau_p \propto N^{-1/2}$ and $I_{peak} \propto N^{3/2}$ in the case of pure Dicke SR, when $T_2$ and photon lifetime are long compared to SF pulse width. Only in the case of short lifetime of photons, the treatment in [7] does recover Dicke's dependence $I_{peak} \propto N^2$ (There is no assumption about short photon lifetime or long decoherence time in [1]). In addition, in [4] , the number and volume of optically excited magnetoexcitons and the critical density for SF are not defined and are not compared to the Mott density of exciton destruction. An unambiguous demonstration would be then under electrical pumping, where initial carrier density can be evaluated precisely. Several claims have been made on SF in bulk GaAs/AlGaAs separate confinement laser diode heterostructure [9], AlGaInAs QWs [10] and InGaN/GaN QWs [11]. However, none of them have demonstrated superlinear peak power growth.

In this letter, we report that we obtained abrupt growth (20%) of peak power of SF pulses with tiny (1%) increase in the amplitude of injected current pulses for electrically pumped InGaN/InGaN QW tandem-cavity lasers with saturable absorber. The peak power growth is accompanied by spectral broadening which is attributed to the narrowing of the pulse width. Cooperative pairing and recombination in highly excited electron–hole plasma is identified as a phase transitions from amplified spontaneous emission. It drastically differs from lasing, revealing (i) an abrupt peak power growth and spectral enlargement, (ii) hyperbolic secant spectral envelope and (iii) red shift of emission. We propose explanation of the observed red shift of SF emission in terms of dipole-dipole coupling and pairing of e-h pairs in an indirect cooperative X-transition.

A separate-confinement heterostructure with three InGaN/GaN quantum wells (QWs) was grown on c-plane GaN free-standing substrate by metal organic vapor phase epitaxy. The wells and barriers are 3.3 and 15 nm thick, respectively. Our structure differs from conventional blue laser diode QW composition [11] and utilizes increased number of QWs and reduced QW width. The waveguiding structure is completed with top *p*-type and bottom *n*-type AlGaN claddings with a *p*-type GaN cap layer [12]. The laser ridge is formed by reactive ion etching with strip width varying from 2 to 3 μm. Index-guided tandem-cavity devices are realized by standard fabrication techniques where the gain and the SA sections are defined by metal contact and with etching of 15 microns wide trenches (Fig. 1a). The tandem cavity of the overall length 800 um incorporates 150 to 240 um long absorber section placed between two gain sections. The individual laser devices were diced and mounted on a heat sink, enabling precise temperature control. At zero absorber bias and pulsed driving of the gain sections, the lasing threshold is 100-150mA.

All measurements are performed at room temperature. The absorber section is negatively biased from a precision voltage source (10 mV accuracy at -20V bias). The gain section of the device is driven from a current source delivering 2 A pulses of the width down to 9 ns. At high absorber bias (-20V) the pulse repetition rate is kept below 5 kHz to avoid absorber breakdown. The injected current pulse amplitude is measured using an inductively coupled current probe. The collimated output beam is sent to a spectrometer of 0.025 nm resolution and an ultrafast photodetector (7 ps response time, 50 GHz bandwidth) whose output is monitored on the sampling scope (70 GHz sampling head). The scope can be triggered either electrically from the pumping pulses monitored by the inductive current probe or optically from the emitted output pulses. For optical triggering, a second photodetector (29 ps/12 GHz) and an optical delay line are incorporated into the setup. Simultaneous observation of the optical spectrum and waveform allows us unambiguously identify the dynamic regimes of emission and tune the number of injected carriers (via current pulse amplitude and its width) as well as the absorber bias so as to produce solitary SF pulses. The probability to generate SF pulses in a pulse-on-demand mode is evaluated from the average repetition rates of the current pulses and the optical triggering events. To verify the small temporal width of generated pulses and confirm recombination of majority of injected carriers, spectrochronograms were acquired using a single-shot streak camera with 2 ps resolution. The pulse intensity incident at the input slit of the camera was strongly attenuated to achieve quoted temporal resolution. The devices were driven by current pulses at repetition rate 25 Hz.

The typical map of dynamic regimes in a tandem InGaN/InGaN cavity is shown in Fig. 1(b) as a function of the absorber bias and amplitude of the current pulse at the gain section. As claimed in [10], superfluorescence in electrically driven laser diode may occur in the transition region from amplified spontaneous emission (ASE, gray domain) to Q-switching lasing (blue domain). Both SF and lasing requires that the gain exceeds losses and both are triggered by spontaneous noise [13]. The difference between emission regimes is imposed by the principle normal mode (solution) of the Maxwell Bloch equations and can be appreciated from the temporal evolution of the process and from its integrated spectrum. Thus, spontaneous emission decays at relaxation time $T_1 \sim 1$ ns of the carrier density $n$. Laser field $E$ decays at the photon lifetime in the cavity. In the case of SF, a macroscopic dipole P builds-up at the characteristic time scale $T_2$ is defined by decoherence processes. This oscillating macroscopic dipole radiates strong electromagnetic pulse. Further difference in behaviour of dynamical variables n, E and P in case of Q-switched lasing and SF emission is displayed by numerical simulations in [11]. Unfortunately, direct measurements of the microscopic dipole magnitude are not possible. Therefore we measure spectral and temporal parameters of emitted pulses in order to distinguish between ASE, SF and lasing emission. In particular, SF emission is expected to reveal superlinear growth of pulse peak power. A fundamental difference also occurs in the state of the e-h system after emission of the light pulse. In the case of Q-switched lasing, the carrier density is just below threshold, so as there should be some ASE observed. In contrast, after a SF burst, the carrier density drops to zero, so no ASE is possible. Using a single-shot spectrochronogram, we examine ASE intensity just after the emission of the main pulse.

SF requires accumulation of high initial carrier density. In a particular GaN device, the estimated carrier density $5\times10^{20}$ cm$^{-3}$ is reached at threshold current 860 mA and absorber bias -20V. The total number of electron-hole pair is $8\times10^{9}$. Interestingly, this is much higher carrier number than previously reported for GaAs [14].

Figures 1 (c) and (d) show the difference between SF (blue data points) and Q-switching (red data points) regimes seen with ultrafast detector and a sampling scope triggered by the

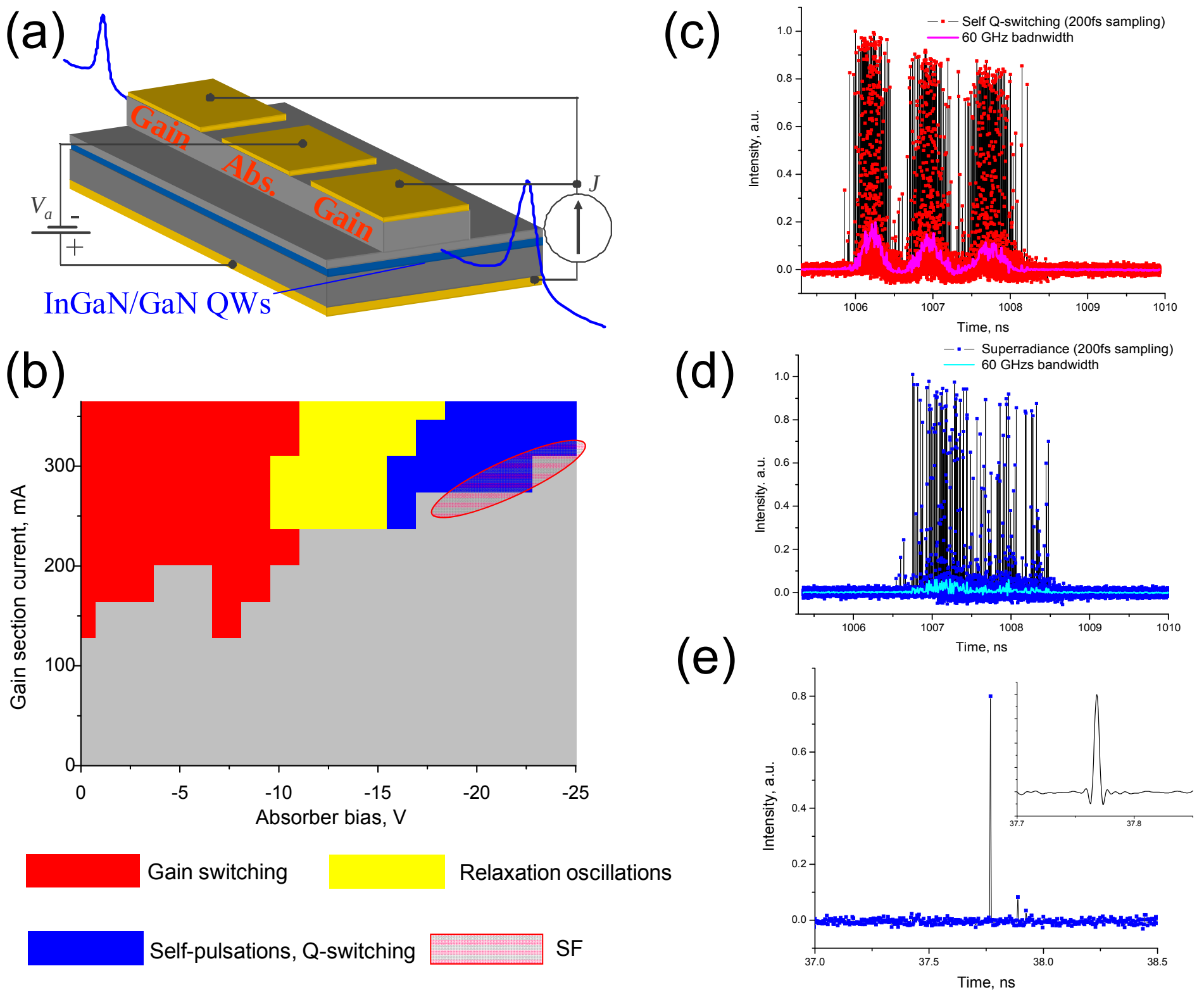

**Figure 1.** (a) Schematic of tandem cavity arrangement and driving conditions; (b) map of lasing regimes; (c) Q-switched lasing, (d) SF and (e) pulse waveforms detected using sampling scope with sampling head 70 GHz and 50 GHz bandwidth detector. In (c) and (d), the scope is triggered by pulses of the pump current. In (e), optical triggering is used. The cyan and magenta curves shows the output of low-pass filter with 60 GHz cut-off applied in data processing.

pump current pulses. (A justification that emission regime in Figs.1 (d) and (e) is SF can be found below, in discussion to Figs.3-6). Narrow-width optical pulses with a high jitter in the SF regime disappear after applying a 60 GHz low- pass filter in the data processing [see cyan curve in Fig. 1(d) in contrast to Fig. 1(c)]. In contrast to this, regular Q-switched pulses do not disappear after such low-pass filter (magenta curve). To confirm that dispersion in the detected sampled SF pulses originates from a pulse jitter (as opposed to broad pulse envelope), we perform same measurements but with the sampling scope triggered at the optical pulse itself. The resulting waveform in Fig. 1(e) shows a single-event trace and attests generation of solitary SF pulse. The width of detected pulse is limited by temporal resolution of the detector. (an example of Q-switched lasing waveform acquired with the optical trigger can be found in Fig. 3(b) )

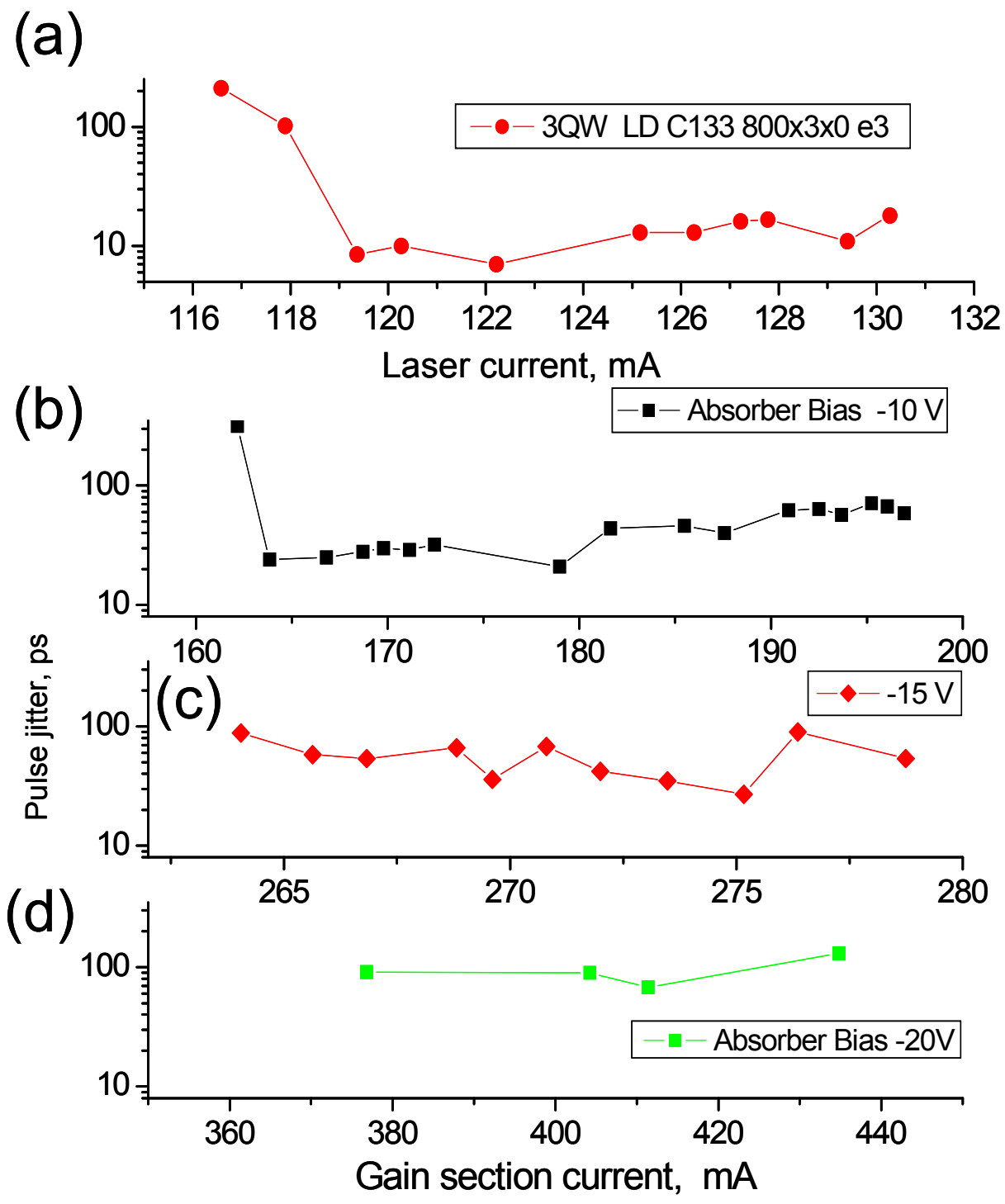

**Figure 2.** Optical pulse jitter in (a) laser with 3 QW active region, in (b) multiple section laser diode operating in gain-switching, and in Q-switching regimes[(c) and (d)]. Absorber biases are -10 V (b) , -15V (c) and -20V (d). The cavity length is 800 µm, and ridge waveguide width is 3 um. In the multiple-section device in (b), (c) and (d), the 100 µm long absorber is placed at the center of the cavity. At -10 V bias and large current, the device operates in gain switching regime. SF is observed at the threshold of Q-switching. From (a) to (d), the threshold current is 115, 160, 262 and 370 mA.

In [11], a special emphasis has been made on the difference in jitter behavior in the lasing regime and SF regime. In particular, it was pointed out that jitter rises abruptly in the SF regime as opposed to lasing. To verify this observation, we perform statistical measurements of the delay time to the pulse emission and estimate pulse jitter as a standard deviation. Our measurements were done both in a single-section laser operating in gain switching regime (Fig. 2(a)) and in a multiple-section device operating in the gain switching (Fig. 2(b)), Q-switching and SR regimes (Fig. 2 (c) and (d)). Both devices have the same cavity length of 800 μm. In contrast to observations reported in Ref. [11], we find that (i) jitter significantly rises at the threshold of a conventional laser (Fig. 2(a)) or multiple section laser operating in gain switching regime at small absorber bias (Note that in Fig 2.(b), the absorber bias is significantly stronger as compare to [11]). At large gain section current and absorber bias of -15 or -20 V, it operates in Q-switching regime. SF occurs only close to threshold (first data point in Fig. 2 (c) and (d)). In our measurements (ii) we don't observe a remarkable difference in the sudden increase of the jitter at the transition from Q-switched lasing to SF regime compare to its dramatically sudden increase at the threshold of the gain-switched lasing.

The difference between SF, ASE and lasing regimes at large absorber bias (-19 to -20 V) is clearly seen from the integrated optical spectrum in Fig. 3(a) and the waveforms detected simultaneously on the sampling scope in Fig. 3(b) (ASE waveform cannot be detected because the peak optical power is well below the noise equivalent power of the detector). Due to low pulse repetition rate (5 kHz) and low duty cycle ($\sim 10^{-5}$), the thermal red shift of the emission wavelength between Q-switched lasing and SF is negligible.

SF is observed in narrow transition region from ASE to stable Q-switched lasing as indicated by the spectral and waveform traces acquired at 760 mA driving current amplitude and -19V absorber bias. Note that the waveform trace in Fig. 3(b) cannot be attributed to possible stochastic behaviour caused by instabilities in the driving current amplitude. Such instabilities would just produce accidental realizations of either lasing or ASE regimes at near lasing threshold operation conditions. They cannot explain the spectral red shift in Fig. 3(a) that was simultaneously observed at 760 mA driving current.

The phase transition to SF regime, which is identified via a spectral red shift in the narrow transition region between ASE to Q-switching lasing, is further examined in Fig. 4(a). Here, the pump current amplitude was varied with the smallest steps enabled by the pulsed current source (0.1 V pulse amplitude increment). The absorber bias was increased to -20V. For comparison, Fig. 4(b) shows the optical spectra acquired at zero voltage bias at absorber.

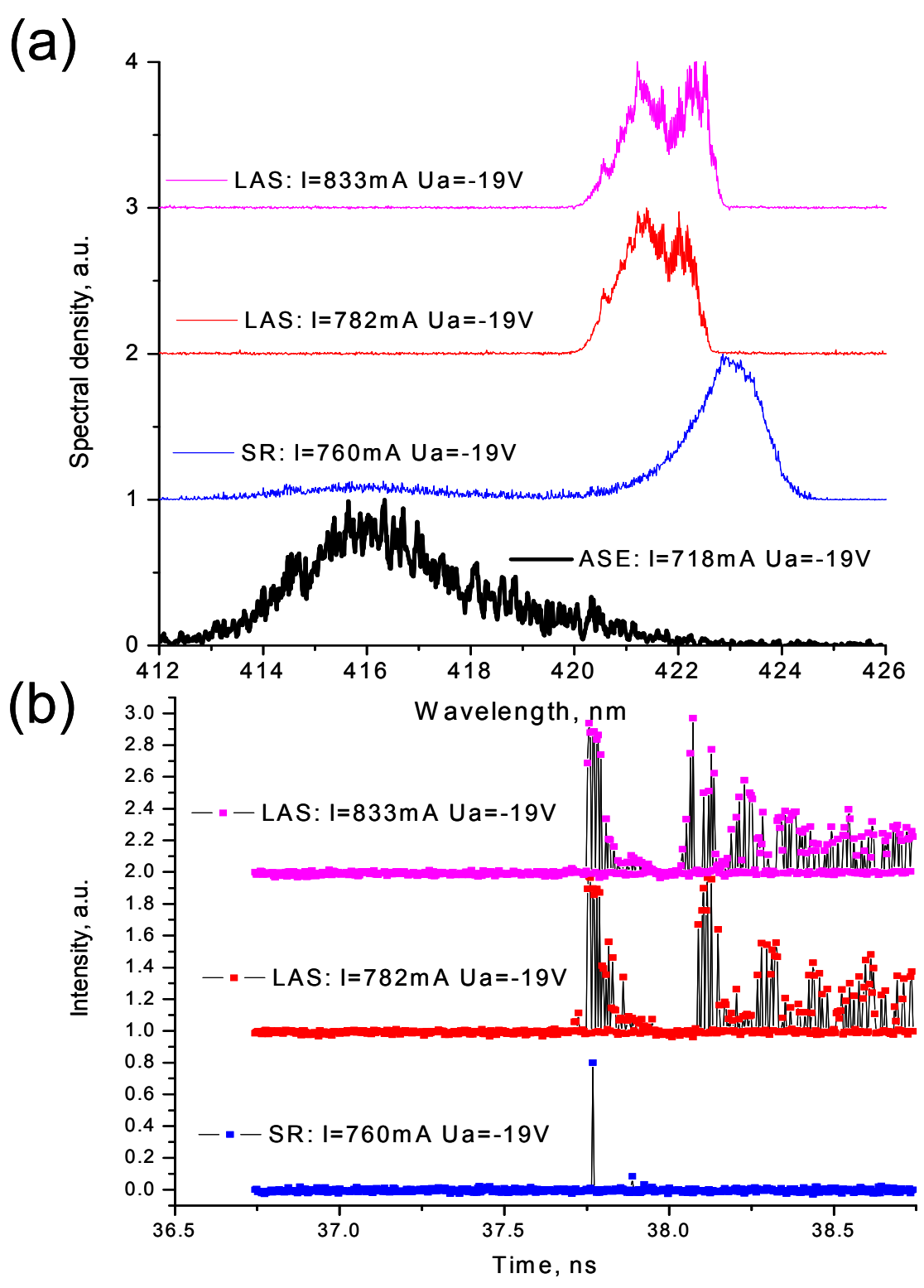

**Figure 3.** Evolution of (a) normalized integrated spectrum and (b) corresponding optical pulse waveform with pump current at large absorber bias ~ -19 V. The cavity length is 800 μm with 240 μm absorber in the center. Driving conditions are indicated in the figure. The gain section is driven by pulses of 12 ns width and repetition rate of 5 kHz. The scope is triggered by the leading edge of the optical pulse.

At zero absorber bias (Fig. 4(b)), the lasing line is red shifted by 0.4 nm (3 meV) from the ASE peak. At high negative absorber bias of -20V (Fig. 4(a)), the threshold current injected into the gain section QW is significantly higher. The ASE at just below threshold exhibits a strong blue shift of 4 nm (30 meV) compared to zero bias due to band filling effect, as the carrier density is about 6 times higher. (Strong carrier screening of the internal built-in field in the gain section QWs is already present at threshold conditions and zero absorber bias and cannot further contribute to the blue shift of the ASE in Figs. 4(a) and 4(b)).

SF, lasing and ASE have clearly distinct spectral features. At high absorber bias, SF at 423 nm is red shifted by 1 nm (7 meV) from the center of the lasing line (at 422 nm) and by 7 nm (50 meV) from ASE peak. Operation in the SF regime at the gain section current 872-879 mA and transition to lasing at 883 mA is evidenced by (i) red shift of SF from lasing and (ii) change of the spectral envelope. The transition is also confirmed by monitoring the pulse waveform on a sampling scope driven by optical trigger (see example in Fig. 3 (b)). The observed SF can not be confused with superluminescence, which is nothing more than ASE in edge emitting cavity with reduced optical feedback from the cavity facets, yielding an increased lasing threshold. The superluminescence spectrum shows a gradual increase of the blue shift of ASE and appearance of the cavity mod ripple with increasing pump current. We also observed such blue shift of ASE (compare the black curves in Figs. 4(b) and 4(a)) . In contrast, transition to SF is accompanied by a sudden red shift of emission.

The band gap energy in InGaN QWs is difficult to define from ASE spectrum as the band edge is affected by inhomogeneous broadening due to QW thickness and In composition variations. Nevertheless, the presented data clearly attests that SF occurs well above the band edge. Thus, at zero absorber bias, ASE and lasing exhibit peak wavelengths in the range 421-422 nm (Fig. 4(c)) with low-energy tail extending down to 426 nm. The SF in Fig. 4(a) is centred at 423 nm, close to the ASE peak at zero absorber bias. Thus, contrary to the statement made in Refs. [9,14, 15], we observe that SF does not occur at the band edge. To further support this observation, in Fig. 4(c) we show modelling results for normalized modal gain (right axis) and net modal gain calculated as a difference between modal gain and absorption weighted by fractional length (left axis). The model calculations utilize isotropic parabolic band approximation and account for inhomogeneous broadening and quantum-confined Stark effect energy shift in the absorber QWs [16]. The results are shown for zero (red curves) and high negative bias (blue curves) of absorber. The peaks of the net modal gain and of the modal gain define, respectively, the wavelength of lasing and ASE. To justify the use of the modal gain spectrum as an approximation to the measured ASE, we examined the high energy tail of the measured ASE with Boltzmann distribution. The numerical fits continuously report temperature values ~100-120 K linearly increasing with the current (and negative absorber bias) in the range of 500-900 mA. This is well below the room-temperature operation conditions of the experiment, attesting that measured spectra are mostly defined by the spectral distribution of the modal gain and not by the spectrum of spontaneous emission (SE), which would exhibit wider high-energy tails [17].

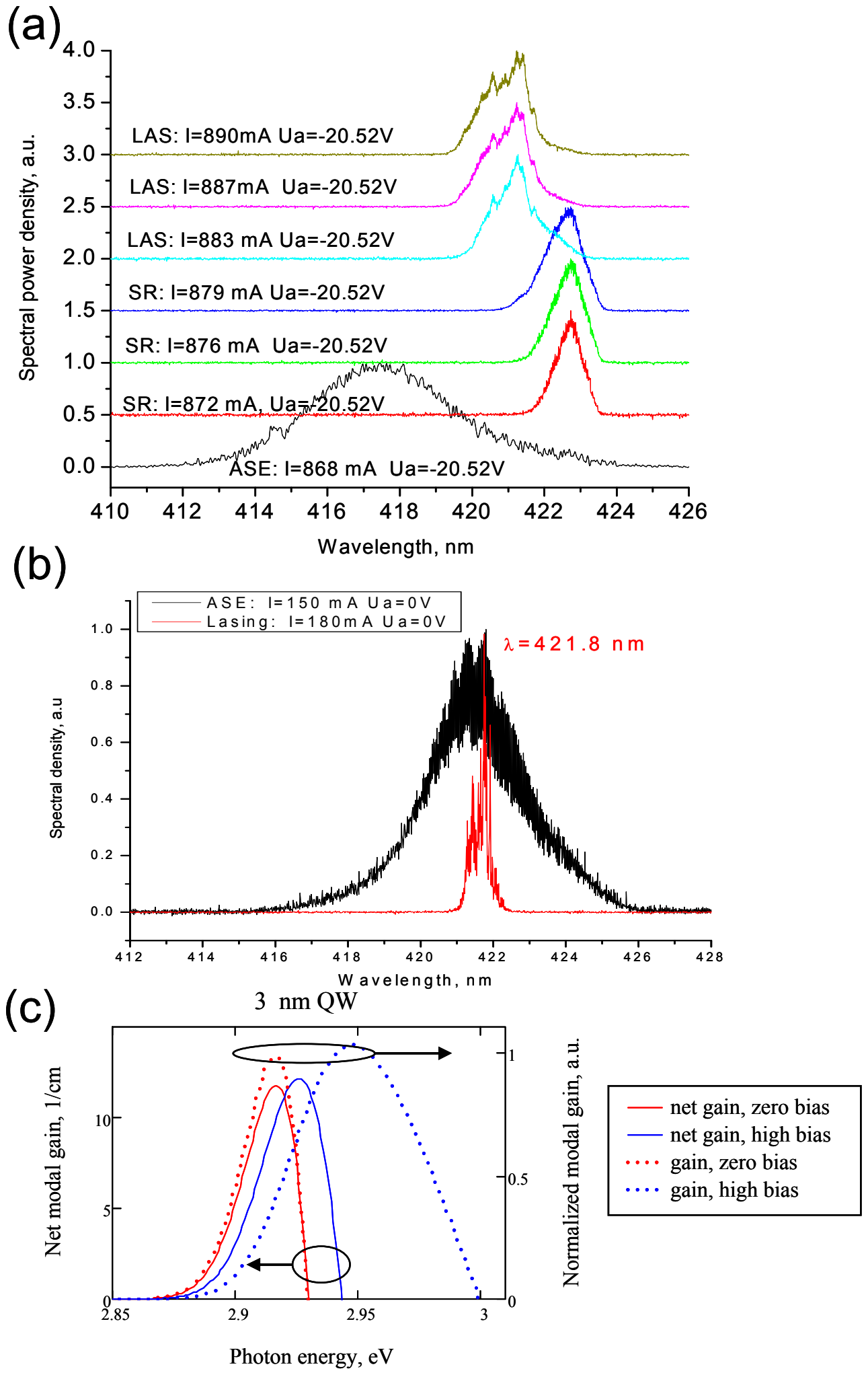

**Figure 4.** Evolution of normalized integrated spectrum with pump current at large absorber bias -20V (a) and zero absorber bias (b). (c) shows model predictions for the net modal gain (solid curves, left axis) and modal gain (dotted curves, right axis) at zero bias (red curves) and high negative bias of absorber.

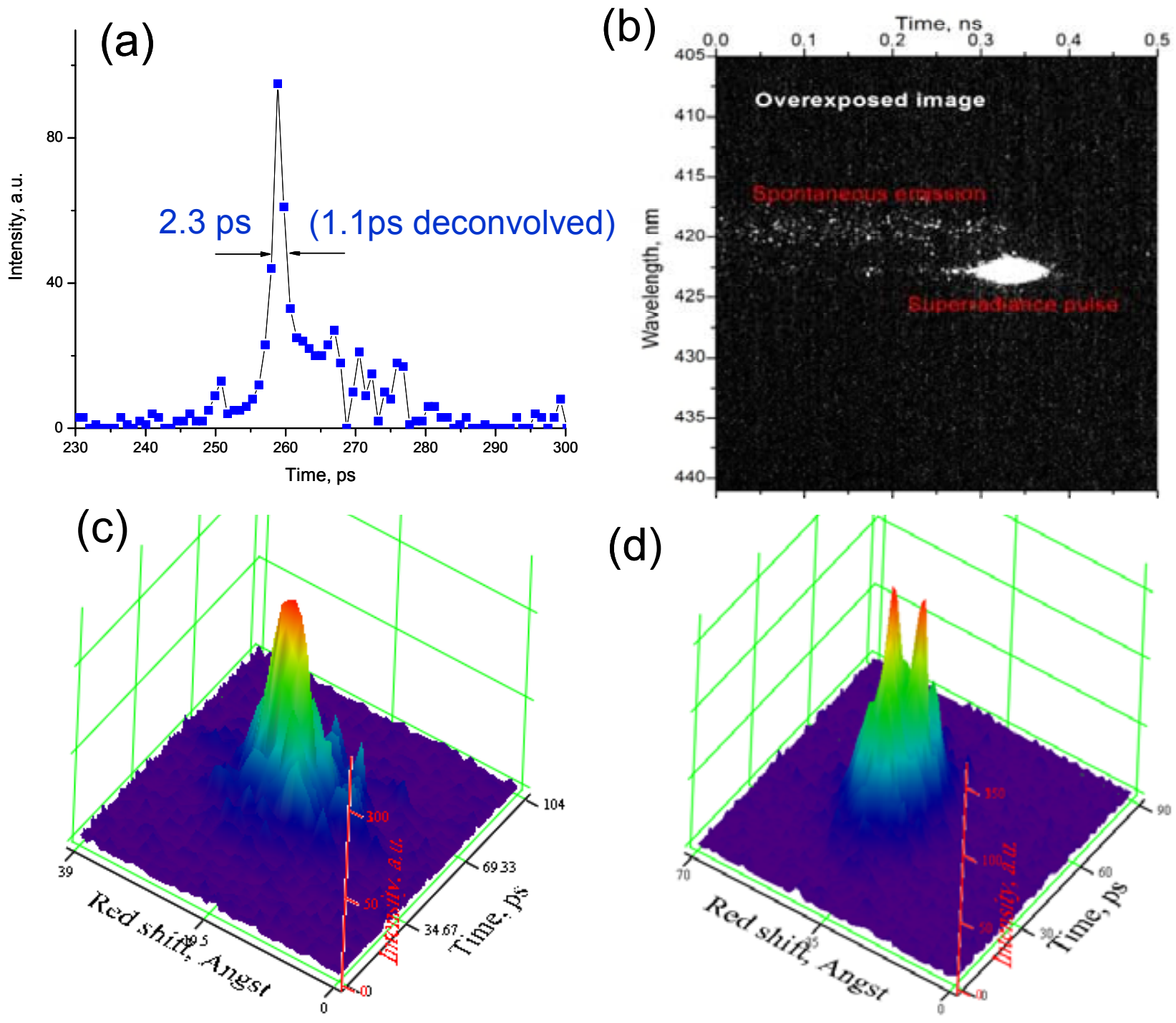

**Figure 5.** SF pulses detected with a 2 ps single-short streak camera. (a) Measured 2.3 ps pulses (1.1 ps deconvolved) in 800 µm cavity with 160 µm central absorber and 3 QW active region driven by 588 mA pulses and absorber bias of -22 V. (b) overexposed spectrochronogram of SF pulse emission in 800 µm cavity with 150 µm central absorber and driven by 417 mA pulses and absorber bias of -23 V. (c) and (d): false color representation of spectrochronogram of SF pulses in 800 um long cavity and 160 um absorber. (c): at 570mA pump current pulses and -22 V absorber bias. (d): at 645 mA current pulses and -24 V absorber bias. The current pulses are of 20 ns width with repetition rate of 25 Hz.

In Fig. 4(c), the spectral features of the calculated modal gain (right axis) qualitatively reproduce the measured ASE spectra. In particular, a large blue-shift of the ASE photon energy occurs due to the band filling effect at high carrier densities. At the same time, the net modal gain maxima at zero and high negative bias of absorber (Fig. 4(c), left axis) do not differ significantly from each other and are close to the modal gain peak at zero bias. (The net gain for the cavity mode is a difference between the modal gain in gain sections and the modal absorption in absorber section). These features qualitatively reproduce experimental spectra in Fig. 4(a) and 4(b) showing nearly the same lasing wavelengths at absorber bias 0V and -19V. SF thus occurs at photon energies close to the lasing photon energy, and hence cannot be attributed to e-h pair condensation at the band edge. However these model simulations do not explain yet why SF is red shifted from the net modal gain maximum.

Figure 5(a) shows one of the *realizations* of SF pulses detected on single-shot streak camera. The deconvolved pulse width is below 1.1 ps under the assumption that the measured pulse width is resolution limited. Overexposed spectrochronogram [Fig 5.(b)] shows that there is no ASE after SF burst, attesting that majority of injected carriers contributed to a single SF pulse. In this particular realization we were able to produce solitary SF pulses in the pulse-on-demand mode. Increasing the width of the pump current pulse, it was possible to reach generation of secondary (lasing) pulses (not shown in the figure). In all cases, we observed a red-shift of the first pulse from the emission spectrum of secondary (lasing) pulses. This red wavelength shift is typical for SF emission, as discussed in relation to Fig. 4. Under very high negative bias of absorber and high pump current, we observe spectral splitting of the SF pulse (compare Fig. 5(c) and (d)). Previously, the doublet line of SF emission has been reported only for GaAs based devices [9]. The doublet line can be attributed to the establishment of oscillatory SF in the presence of (i) very high optical gain and (ii) long propagation distance [18]. The pulse envelope decays at the time scale of the effective dephasing time $T_2$. The period of these oscillations estimated from the spectral splitting of the doublet line in Fig. 5(d) is about 0.4 ps, which agrees with the expected time scale set by the dephasing time. The temporal evolution of oscillatory SF cannot be resolved in our measurements.

The oscillatory SF at sub-picosecond time scale and formation of spectral doublet is drastically different from periodic train of (lasing) pulses at the cavity roundtrip time reported in wide-QW InGaN based devices with 400 um cavity length [11]. This last one decays during several cavity roundtrips. In contrast, the oscillatory SF in Fig.5 (d) exhibits pulse envelope shorter than the cavity roundtrip time (19 ps). As shown by several independent numerical simulations [11,19], in the SF regime almost all energy stored in the inverted medium are released during a single passage in the cavity of the building-up SF pulse. Such dynamics can be observed in Fig. 3 and 5.

An estimate for the lower boundary of the (spectrally limited) SF pulsewidth is obtained from the optical spectrum, assuming pulse envelope $E(t) \propto \mathrm{sech}(t/\tau_p)$ (see e.g. [20]) with $\tau_c = 2\tau_p$ being the pulse width. The optical spectrum is defined by its Fourier transform $E(\omega) \propto \mathrm{sech}\lfloor \pi\tau_p(\omega - \omega_0)/2 \rfloor$ (see e.g. [21]). In Fig. 6(c), the numerical fit with $E(\omega)^2$ of measured SF spectra at I = 883, 887 and 890 mA from Fig. 4(a) reports the pulse width in the range of 0.2 ps (an example of the fit is shown in the inset). The actual pulse width is between the one limited by the streak camera resolution (1.1 ps) and the spectrally limited one. The peak power in Fig. 6(d) was estimated from the measured average power [Fig. 6(a)], taking away the ASE power measured at just below the SF threshold, using spectrally limited pulsewidth and accounting for the limited probability to generate SF pulses. (Not each applied current pulse produces SF emission. This circumstance has to be accounted for in future applications of SF emitters.) Interestingly, the spectrally limited pulse width ~0.2 ps and peak power ~200 W are comparable to the best reported so far values for GaAs devices [9,15]. These agree with the numerical simulations and analytic calculations from Ref. [20] for 800 um long devices.

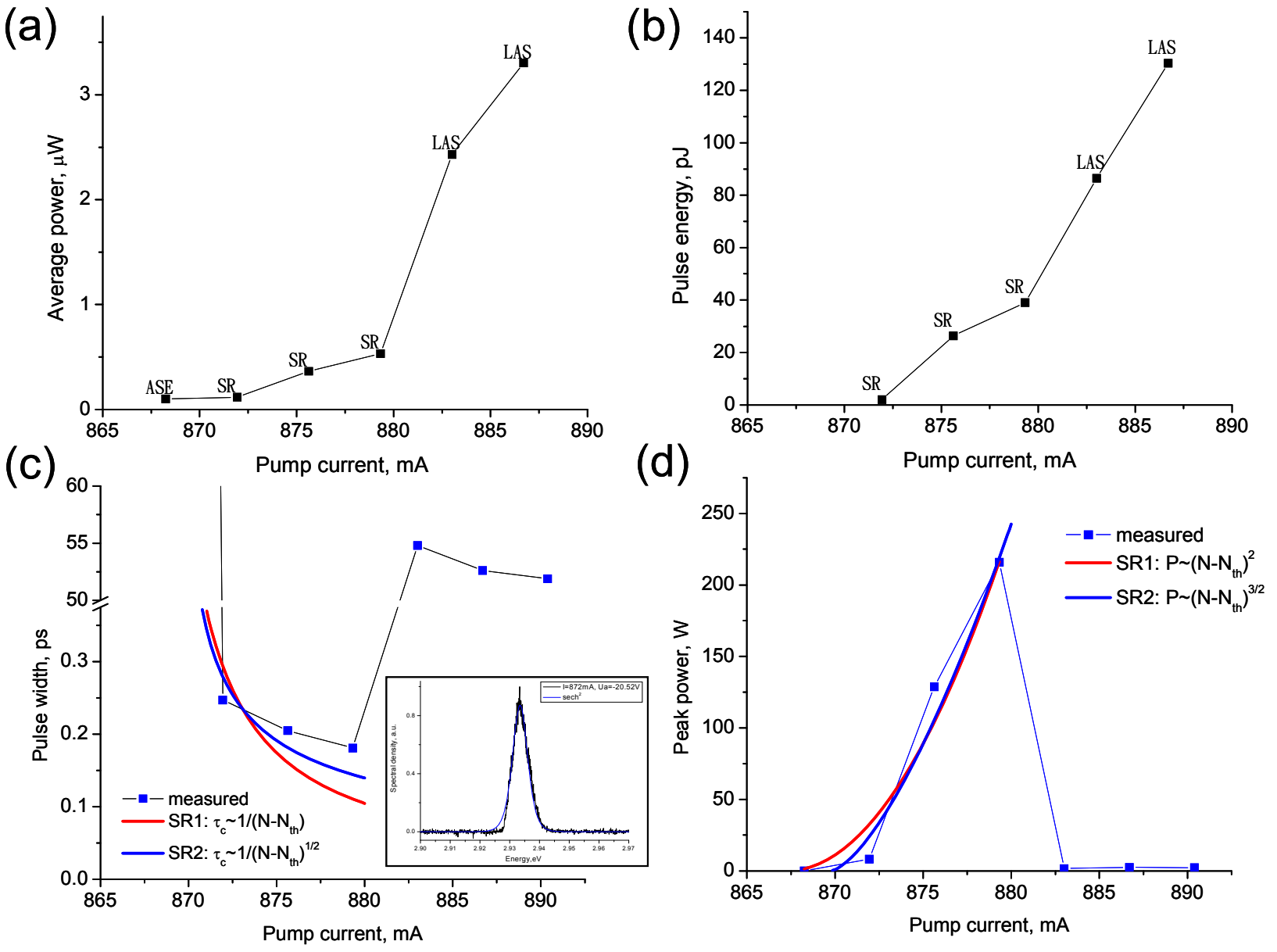

**Figure 6.** (a) Average power and (b) pulse energy. (c) Measured (points) and numerically fitted (blue and red curves) spectrally limited pulse width and (d) peak power vs current in measurements of Fig. 4. Red and blue curves distinguish type-1 SF and type-II SF. The inset in (c) shows an example of measured (black curve) and modeled (blue curve) SF spectrum.

In [11,20], it was shown that type-I SF with Dicke's quadratic growth of the peak $I_{peak} \propto (N - N_{cr})^2$ and pulse width $\tau_c \sim (N - N_{cr})^{-1}$, with $N_{cr}$ being SF threshold, can be reached in short samples. In long samples as we use here, type-II SF would be observed with a gentler rise of the peak power $I_{peak} \propto (N - N_{cr})^{3/2}$ and pulse width $\tau_c \sim (N - N_{cr})^{-1/2}$ due to formation of coherence domains in the active region. In Figs. 6(c) and 6(d), we present the results of simultaneous numerical fits utilizing type-I SF (red curve) and type-II SF (blue curves) models. Note that a possible linear growth approximation for the peak power $I_{peak} \propto (N - N_{cr})$ in Fig. 6(d), as in the usual lasing regime, does not match the data in Fig. 6(a). Indeed, with pulse energy $\mathrm{E} = I_{peak}\tau_c \propto (N - N_{cr})$ [Fig.6 (b)], the pulse width would remain constant in this approximation. In Fig. 6, the numerical fits assuming type –II SF exhibit better agreement with experimental data for the pulse width and peak power. This is an expected result as the sample is 800 um long. One may conclude that type-II SF has been observed. The *measured* pulse width is not larger than 1.1 ps and estimated peak power is at least 70 W.

It now remains to explain the *small* red shift of SF emission from the peak of the gain curve. In [9,15], condensation of e-h pairs is predicted to occur few meV above the band edge, but in our experiment we observed it close to lasing emission which is far from the band edge. In [6], it was *stated* that SF should occur at about half of the Fermi energy and that it is blue-shifted form the gain peak. Both *hypotheses* are in contradiction to our measurements.

In [11,20], it has been shown that single-mode SF in long sample is governed by Ginzburg Landau –like (or Gross-Pitaevskii -like) equation for polaritons. The squared magnitude of macroscopic dipole $\mathrm{p}^2$, which is expressed in number of coherence excitations, defines scattering energy and fulfils the Bloch vector conservation $\mathrm{p}(t)^2 + (N(t) - N_{cr})^2 = (N_0 - N_{cr})^2$ with $N_0$ being the initial carrier number. To explain the red shift of SF emission, we will relate the output radiation intensity to the spectrum $\mathrm{p}^2$ of pair-wise dipole-dipole correlations.

For a few paragraphs, we temporarily ignore the intraband dispersion of electrons and holes as well as the spin degeneracy and make use of symmetric Dicke states $|J,M\rangle$ for a coherent ensemble of two-level quantum systems. In application to the system of e-h pairs, $2J+1=2(N_0-N_{cr})$ is the number of initially excited pairs above the critical density and $2M$ is the excess of e-h pairs in $|J,M\rangle$ state that have not recombined by the time *t*. The functions $|J,M\rangle$ constitute a basis of 2*J*+1 dimensional irreducible representation (IR) of $SO_3$ group [22]. The Bloch vector conservation follows from the operator relationship $\hat{J}^2=\hat{J}_z^2-\hat{J}_z+\hat{J}_+\hat{J}_-$ with the excitation number of macroscopic coherences being defined by the product of the lowering and rising operators in the ensemble $\mathrm{p}(t)^2=\langle\hat{J}_+\hat{J}_-\rangle=\sum_{M=-J}^{J}\langle JM|\hat{J}_+\hat{J}_-|JM\rangle P_M(t)$, where $\langle JM|\hat{J}_+\hat{J}_-|JM\rangle=(J+M)(J-M+1)$ and $P_M(t)$ is the probability to find the system at time *t* in a pure state $|J,M\rangle$, $\sum_{M=-J}^{J}P_M(t)=1$. In case of large carrier numbers, other terms of this operator relationship yield $\langle\hat{J}^2\rangle=J(J+1)\approx(N_0-N_{cr})^2$ and $\langle\hat{J}_z^2-\hat{J}_z\rangle=\sum_M M(M-1)P_M(t)\approx(N(t)-N_{cr})^2$.

The phase of periodic Bloch functions in the conduction and valence bands can be selected in such a way that the matrix element for a direct transition in dipole radiation approximation $\langle h|\hat{d}^{(i)}|e\rangle=\mu$ is a real number (see e.g. [23]). In this case, for a solitary e-h pair, $\hat{d}^{(i)}=2\mu\hat{J}_x^{(i)}=\mu(\hat{J}_+^{(i)}+\hat{J}_-^{(i)})=\hat{d}_+^{(i)}+\hat{d}_-^{(i)}$, where $\hat{J}_\alpha^{(i)}=\frac{1}{2}\hat{\sigma}_\alpha$, $\hat{\sigma}_\alpha$ is the Pauli matrices with indexes $\alpha=\{x,y,z\}$ and the upper index *i* enumerates e-h pairs. The direct product of the wavefunctions of *two* e-h pairs constitutes a basis of a four-dimensional representation of $SO_3$. It can be reduced to a symmetric three-dimensional subspace of $SO_3$ spanned by the basis set $|1,1\rangle^{(i,j)}=|e^{(i)},e^{(j)}\rangle$, $|1,0\rangle^{(i,j)}=\frac{1}{\sqrt{2}}|e^{(i)},h^{(j)}\rangle+\frac{1}{\sqrt{2}}|h^{(i)},e^{(j)}\rangle$, $|1,-1\rangle^{(i,j)}=|h^{(i)},h^{(j)}\rangle$ and antisymmetric singlet state $|0,0\rangle^{(i,j)}=\frac{1}{\sqrt{2}}|e^{(i)},h^{(j)}\rangle-\frac{1}{\sqrt{2}}|h^{(i)},e^{(j)}\rangle$. The states $|1,m\rangle^{(i,j)}$ and $|0,0\rangle^{(i,j)}$ are the eigen states of the z-component of the angular momentum operator $\hat{\mathbf{J}}^{(i,j)}=\hat{\mathbf{J}}^{(i)}+\hat{\mathbf{J}}^{(j)}$. However, the antisymmetric singlet state remains a dark state during cooperative SF emission [24], and it can be excluded from consideration. Considering a total angular momentum of the system $\hat{\mathbf{J}}=\sum_{i=1}^{2J}\hat{\mathbf{J}}^{(i)}$ and macroscopic polarization $\hat{d}=\mu(\hat{J}_++\hat{J}_-)=\sum_{i=1}^{2J}\hat{d}^{(i)}$, one can show that [24] (i) the mean radiated intensity is defined by macroscopic polarization $I(t)\propto\mu^2\mathrm{p}(t)^2=\langle\hat{d}_+\hat{d}_-\rangle$. In addition, (ii) macroscopic polarization can be resolved in two main terms, originating from pair-wise dipole-dipole cross-correlations and self-correlations respectively, yielding $\left\langle \hat{d}_+\hat{d}_- \right\rangle = \sum_{i\neq j}\left\langle \hat{d}_+^{(i)}\hat{d}_-^{(j)} \right\rangle +$ $+\mu^2\sum_{M=-J}^{J}(J+M)P_M(t)$. The contribution from cross-correlation term is by a factor of 2$J$ higher due to the number of terms involved in the sum. Respectively, the spectrum of $I(t)$ is mainly defined by pair-wise dipole-dipole cross-correlations. In order to calculate it, this term will be represented in the basis of periodic Bloch functions $\left|e\right\rangle$ and $\left|h\right\rangle$.

The ensemble-average of the cross-correlations between two e-h pairs $\left\langle \hat{d}_+^{(i)}\hat{d}_-^{(j)} \right\rangle = \sum_{M=-J}^{J}\left\langle JM\right|\hat{d}_+^{(i)}\hat{d}_-^{(j)}\left|JM\right\rangle P_M(t) = \mu^2\sum_{M=-J}^{J}\left\langle JM\right|\hat{J}_+^{(i)}\hat{J}_-^{(j)}\left|JM\right\rangle P_M(t)$ is calculated here by reducing the matrix element to the 3-dimentional IR subspace of the basis $\left|1,m\right\rangle^{(i,j)}$. With expansion $\left|JM\right\rangle = \sum_{m=-1}^{1}\left|J-1,M-m\right\rangle\otimes\left|1,m\right\rangle^{(i,j)}\left\langle J-1,M-m,1,m\middle|J,M\right\rangle$, which utilizes Clebsch–Gordan coefficients, we find that $\left\langle JM\right|\hat{J}_+^{(i)}\hat{J}_-^{(j)}\left|JM\right\rangle$ has only one nonzero term originating from reduced matrix element ${}^{(i,j)}\left\langle 1,0\right|\hat{J}_+^{(i)}\hat{J}_-^{(j)}\left|1,0\right\rangle^{(i,j)} = \frac{1}{2}$. Finally, substituting the symmetric combinations of $\left|e^{(i,j)}\right\rangle$ and $\left|h^{(i,j)}\right\rangle$ wavefunctions for $\left|1,m\right\rangle^{(i,j)}$, noting that operators $\hat{d}^{(i)}$ and $\hat{d}^{(j)}$ commute and that $\left(\hat{d}_+^{(i)}\right)^+ = \hat{d}_-^{(i)}$, we obtain the SF intensity

$$I_{SF}(t) \propto \sum_M P_M(t)\left\{\frac{J^2-M^2}{2J(2J-1)}\sum_{i\neq j}\left\langle h^{(i)}h^{(j)}\right|\hat{d}_-^{(i)}\hat{d}_-^{(j)}\left|e^{(i)}e^{(j)}\right\rangle + \mu^2(J+M)\right\} \quad (1)$$

In the case of large carrier numbers, the main contribution in (1) is due to pair-wise cross-correlations (the first term in the curly brackets) as it has $2J(2J\text{-}1)$ summands, with each being of the magnitude $\sim\mu^2$. If the intraband dispersion and energy conservation are taken into account in (1), the radiation spectrum and the excitation spectrum of macroscopic coherences at the initial stage of SF emission are defined by squared occupation numbers in the bands

$$\mathrm{I}_{\mathrm{SF}} \propto \mu^2\mathrm{p}^2 \approx \mu^2\sum_{j\neq j} f_c^{(i)}f_c^{(j)}(1-f_v^{(i)})(1-f_v^{(j)})\delta(E_c^{(i)}+E_c^{(j)}-E_v^{(i)}-E_v^{(j)}-2\hbar\omega)\delta_{\mathbf{k}_e^{(i)}+\mathbf{k}_e^{(j)},\mathbf{k}_h^{(i)}+\mathbf{k}_h^{(j)}}, \quad (2)$$

where $f_c^{(i)}$ ($f_v^{(j)}$) is the electron occupation number in conduction (valence) band at $\mathbf{k}_e^{(i)}$ ($\mathbf{k}_h^{(j)}$) in the first Brillouin zone. Obviously, (2) has different spectral features from the linear optical gain in the lasing regime $\mathrm{g}_{\mathrm{LAS}} \propto \mu^2\sum_i(f_c^{(i)}-f_v^{(i)})\delta(E_c^{(i)}-E_v^{(j)}-\hbar\omega)\delta_{\mathbf{k}_e^{(i)},\mathbf{k}_h^{(i)}}$.

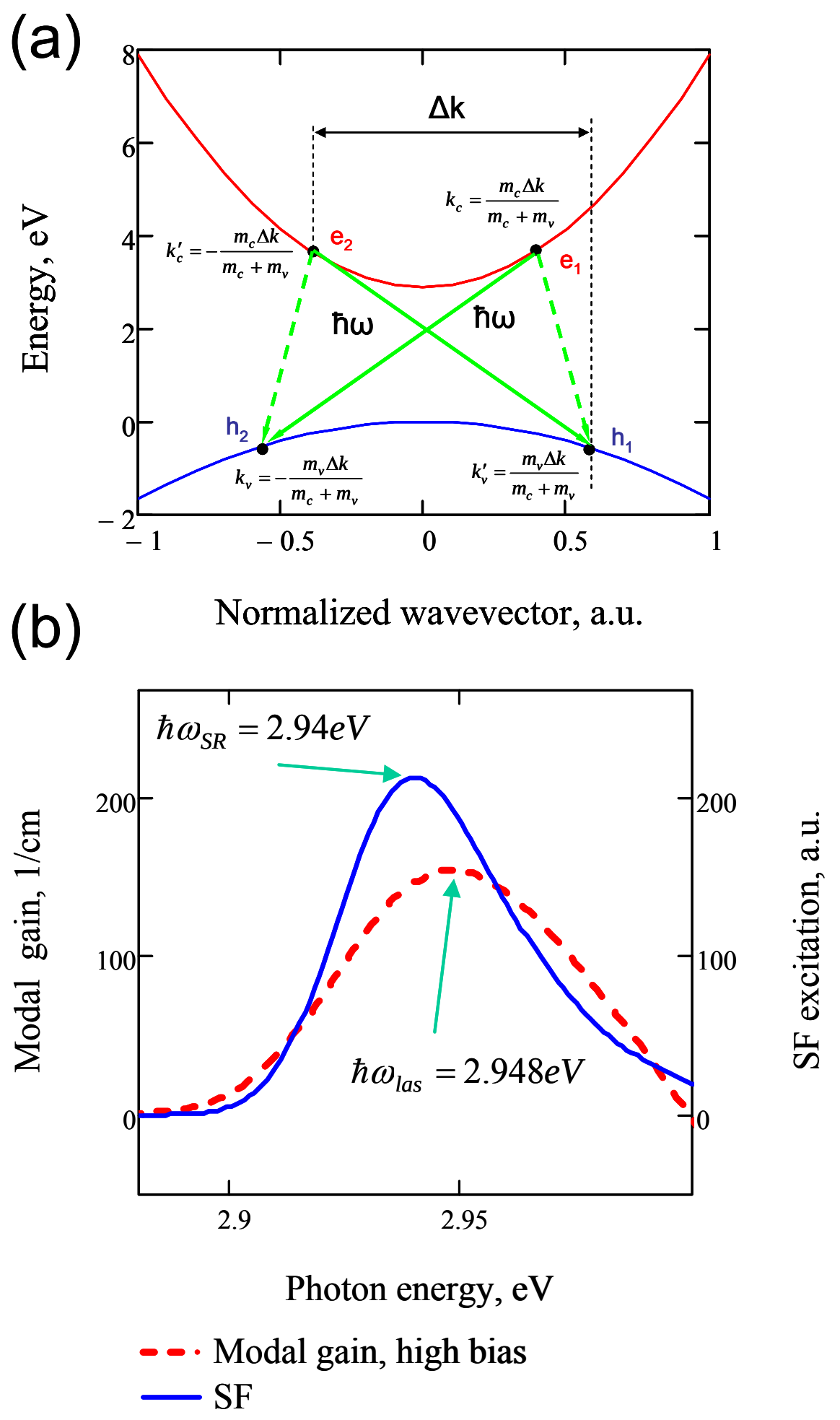

**Figure 7.** (a): Energy diagram and cooperative X transition between $e_1$-$h_1$ and $e_2$-$h_2$ pairs. Green solid arrows and dashed arrows indicate possibility of entanglement at $k_0$=0. (b): normalized SF emission spectrum at initial stage (blue solid curve) for $k_0$=0 and modal gain at high carrier density (red dashed curve).

Interestingly, without lattice phonons, the linear optical gain assumes only a direct transition at $\mathbf{k}_e^{(i)} = \mathbf{k}_h^{(i)}$. However, SF emission admits recombination of two indirect e-h pairs in a cooperative X transition with emission of two photons at same energy Fig. 7 (a) shows such two e-h pairs at different locations on the dispersion curves. The energies and momentum conservation in (2) requires that repulsion Δk and center of mass momentum $k_0$ fulfill the dispersion equation $\hbar\omega = E_g + \hbar^2 \Delta k^2 / 2(m_c + m_v) + \hbar^2 k_0^2 / 2\mu_{ex}$ with $\mu_{ex} = m_c m_v / (m_c + m_v)$ being the reduced exciton mass (Fig. 7 (a) shows the case $k_0$=0). The notion of indirect X transition will allow elaborating details of the mechanism of cooperative recombination of majority of carriers with emission of photons at essentially same energy.

Normalized spectral shapes of the linear modal gain and SF emission spectrum at initial stage (2) in the presence of inhomogeneous broadening are plotted in Fig. 7(b). The peak of SF emission is red shifted by 8 meV from the modal gain. Thus, the proposed concept of indirect cooperative X transition explains the red shift of the SF emission from the lasing wavelength. This is not an unexpected result. At carrier density $5\times10^{20}$ $cm^{-3}$, the mean inter-particle distance is ~1 nm, much smaller than the wavelength. So we will expect strong dipole-dipole correlations between e-h pairs. Note the similarities between proposed concept of indirect cooperative X transition in semiconductors and previously evoked transition diagram for collective annihilation of a magnetized electron-hole plasma in Ref. [25].

As a conclusive remark, we note that X transition at $k_0 = 0$ and $\mathbf{k}_e^{(i)} = -\mathbf{k}_e^{(j)}$, $\mathbf{k}_h^{(i)} = -\mathbf{k}_h^{(j)}$ may lead to entanglement as depicted in Fig.7(a), possibility of which was evoked in Ref. [26].

This research is supported by the EC Seventh Framework Programme FP7/2007–2013 under the Grant Agreement n 238556 ( FEMTOBLUE).